\documentclass[10pt,fleqn,twocolumn]{article}
\usepackage{multirow}
\usepackage{graphicx}
\usepackage{amsmath,amssymb}
\usepackage{bm}
\usepackage{cite}
\usepackage{notoccite}
\usepackage[blocks,auth-sc,affil-sl]{authblk}
\usepackage[top=0.75in, bottom=0.75in, left=0.75in, right=0.75in, dvips]{geometry}

\def\d{\mathrm{d}}
\usepackage{slashed}
\usepackage{subfigure}
\def\d{\mathrm{d}}
\def\beq{\begin{equation}}
\def\eeq{\end{equation}}
\def\bea{\begin{eqnarray}}
\def\eea{\end{eqnarray}}

\def\nnb{\nonumber}
\def\ga{\left(}
\def\dr{\right)}

\def\nnb{\nonumber}

\def\ba{\begin{array}}
	\def\ea{\end{array}}

\begin{document}

\title{The application of light front holographic QCD to B physics}

\author{M. R. Ahmady\thanks{Speaker. E-mail: \texttt{mahmady@mta.ca}.}}

\affil{Department of Physics, Mount Allison University, Sackville, N-B. E46 1E6, Canada.}
\author{S. Lord}

\affil{D\'epartement de Math\'ematiques et Statistique, Universit\'e de Moncton,
	Moncton, N-B. E1A 3E9, Canada.}
\author{R. Sandapen}
\affil{D\'epartement de Physique et d'Astronomie, Universit\'e de Moncton,
	Moncton, N-B. E1A 3E9, Canada
	\& \\
	Department of Physics, Mount Allison University, Sackville, N-B. E46 1E6, Canada.}

\maketitle

\begin{abstract} 
 Light front Holography is a formalism developed by Brodsky and de T\'{e}ramond in which analytic forms for the hadronic bound state wavefunctions can be obtained.  We have used the holographic light front wavefunctions thus obtained in order to calculate the distribution amplitudes of the light vector mesons $\rho$ and $K^*$. As a result, we are able to calculate the form factors for $B\to \rho , K^*$ transitions as well as $\Lambda_{\rm QCD}/m_b$ contributions in radiative $B_{(s)}\to (\rho ,K^*)\gamma$ decays.  We compare our predictions to the available experimental data.
\end{abstract}


\section{Introduction}
Flavor Changing Neutral Currents (FCNC) are excellent probes to the Standard Model (SM) and beyond. In particular, the $b\to(s,d) \gamma$ and $b\to (s,d) \ell^+\ell^-$ transitions, where $\ell =e,\, \mu ,\,\tau$, are most important for the extraction of  the Cabibbo-Kobayashi-Maskawa (CKM) quark mixing matrix elements as well as for the search of New Physics (NP) signals. The experimental measurements for the $b\to d \gamma$ transistion are currently available  for exclusive radiative $B$ decay to a $\rho$ meson, i.e. $B \to \rho \gamma$.  Both exclusive $B\to K^{(*)}$ and inclusive $B\to X_s$ data are available for $b\to s\gamma$ and $b\to s\ell^+\ell^-$.  For a review of radiative $B$ decays, we refer to \cite{Hurth:2010tk}. 

The theory of exclusive decays is complicated by their sensitivity to non-perturbative physics. The standard\footnote{Alternative frameworks can be found in reference \cite{Lu:2005yz,Ali:2001ez}.} theoretical framework for these decays is QCD factorization (QCDF)\cite{Bosch:2001gv,Beneke:2001at}  which states that,  to leading power accuracy in the heavy quark limit, the decay amplitude factorizes into perturbatively computable kernels and non-perturbative objects namely the $B\to V$ transition form factors\footnote{V stands for vector mesons.}, the meson decay constants and the leading twist Distribution Amplitudes (DAs) of the mesons. Traditionally the DAs for the vector meson are obtained from QCD Sum Rules (SR)\cite{Ball:1996tb,Ball:1998fj,Ball:1998ff}. The numerical values of the transition form factor and the tensor decay constant of the vector meson are  obtained from light-cone Sum Rules (LCSR) or lattice QCD. The predictive power of QCDF is limited by the uncertainties associated with these non-perturbative quantities and also by power corrections to the leading  amplitude \cite{Antonelli:2009ws}. The computation of the power corrections is often problematic  due to the appearance of end-point divergences in convolution integrals that contribute to the decay amplitude \cite{Antonelli:2009ws,Pecjak:2008gv}. 

In this work, we use the holographic anti-de Sitter quantum chromodynamics (AdS/QCD) prediction for the light front wavefunction \cite{Brodsky,physicsreport} of the vector mesons to derive the DAs for $\rho$ and $K^*$.  We then use these DAs to predict the branching ratio for the decays $\bar B^\circ\rightarrow \rho^\circ \gamma$ and $\bar B^\circ_s\rightarrow \rho^\circ \gamma$ beyond leading power accuracy in the heavy quark limit.  We also predict the isospin asymmetry as well as the branching ratio for the decay $B \to K^* \gamma$ within QCD factorization using AdS/QCD holographic DAs for the $K^*$ meson. 

Finally, we use LCSR  with AdS/QCD DAs for the $\rho$ and $K^*$ vector mesons to predict the form factors that govern the leading amplitudes of  the rare radiative $B \to (\rho , \, K^*)$ decays as well as the semileptonic decay $B \to \rho l \nu$. We also compute the total width (in units of $|V_{ub}|^2$) for the semileptonic decay $B \to \rho l \nu$ as well as ratios of partial decay widths which are independent of $|V_{ub}|$. Finally, we use our form factors to predict the differential and total branching fraction of the rare dileptonic decay $B \to K^* \mu^+ \mu^-$ which we compare to the recent LHCb data.


\section{QCD Factorization}
At low energies available in B meson decays, one can integrate out the heavier degrees of freedom like $W^\pm$ and $Z^0$ gauge bosons and top quark and work with an effective theory.  The effective Hamiltonian for $b\to s(\gamma ,\, \ell^+\ell^-)$ transitions can be written as:
{\small
	\begin{eqnarray}\label{heff}
	{\cal H}_{eff}=\frac{G_F}{\sqrt{2}}\sum_{p=u,c}V^*_{ps}V_{pb}
	\left[ C_1 Q^p_1 + C_2 Q^p_2 +\sum_{i=3,\ldots ,10} C_i Q_i\right],
	\end{eqnarray}
}
where the operators $Q_i$ with $i=1,\ldots ,10$ are defined as:
{\small
	\begin{eqnarray}\label{q1def}
	Q^p_1 &=& (\bar sp)_{V-A}(\bar pb)_{V-A} \;\;\;
	Q^p_2 = (\bar s_i p_j)_{V-A}(\bar p_j b_i)_{V-A} \nonumber\\
	Q_3 &=& (\bar sb)_{V-A} \sum_q (\bar qq)_{V-A} \;\;\;
	Q_4 =   (\bar s_i b_j)_{V-A} \sum_q (\bar q_j q_i)_{V-A} \nonumber\\
	Q_5 &=& (\bar sb)_{V-A} \sum_q (\bar qq)_{V+A} \;\;\;
	Q_6 = (\bar s_i b_j)_{V-A} \sum_q (\bar q_j q_i)_{V+A} \nonumber\\
	\label{q7def}
	Q_7 &=& \frac{e}{8\pi^2}m_b\,
	\bar s\sigma^{\mu\nu}(1+\gamma_5)b\, F_{\mu\nu}\nonumber\\
	\label{q8def}
	Q_8 &=& \frac{g}{8\pi^2}m_b\,
	\bar s_i\sigma^{\mu\nu}(1+\gamma_5)T^a_{ij} b_j\, G^a_{\mu\nu}\nonumber\\
	Q_9 &=& \frac{e^2}{8\pi^2}\,
	\bar s\gamma^\mu(1-\gamma_5)b\, \bar{\ell}\gamma^\mu\ell\nonumber \\
	Q_{10} &=& \frac{e^2}{8\pi^2}\,
	\bar s\gamma^\mu(1-\gamma_5)b\, \bar{\ell}\gamma^\mu\gamma_5\ell\nonumber
	\end{eqnarray}
}
For $b\to d(\gamma ,\ell^+\ell^- )$ transitions, $s$ should be replaced with $d$ in the above equations.  $C_i$ are the Wilson Coefficients which are evaluated perturbatively and their numerical values to next-to-leading-order are given in \cite{Ball:2006eu,Buchalla:1995vs}\footnote{Note that $C_{1,2}$(here)$=C_{2,1}$(reference \cite{Ball:2006eu}).} and the numerical values of  CKM matrix elements are given in reference \cite{Beringer:1900zz}. 

At leading power in $\Lambda_{\rm QCD}/m_b$ accuracy in the heavy quark limit and to all orders in the strong coupling $\alpha_s$,  the matrix element of an operator $Q_i$ factorizes as \cite{Bosch:2001gv,beneke}
{\small
	\begin{eqnarray}
	&&\langle (\rho , K^*) ( P,e_T)| Q_i | \bar{B} \rangle \\&& = \underbrace{ F^{B\rightarrow \rho ,K^*} T_i^I}_{\rm spectator\; terms} 
	+ \underbrace{\int_0^1 d z \; d \xi \; \Phi_B(\xi ) T_i^{II}(\xi ,z)  \phi^{\perp}_{\rho ,K^*} (z)}_{\rm nonspectator\; terms}\; ,\nonumber
	\end{eqnarray}}
i.e. into perturbatively computable  hard-scattering kernels $T_i^I$  and $T_i^{II}$ and three non-perturbative quantities namely the transition form factor $F^{B\rightarrow \rho ,K^*}$, the leading twist DA of the $B$ meson, $\Phi_B(\zeta) $, and the twist-$2$ DA of $\rho$ or $K^*$ meson, $\phi_{\rho ,K^*}^{\perp} (z)$. In the following sections, we first calculate the DAs for $\rho$ and $K^*$ mesons using the light front holographic wavefunction and then apply the LCSR method to derive the transition form factors for $B\to\rho\, , K^*$.


\section{Distribution Amplitudes}
We work with light cone coordinates, i.e. $x^\mu =(x^+, x^-,x_\perp )$, where $x^\pm =x^0\pm x^3$ and $x_\perp$ is equal to ($x_1$, $x_2$).  At equal light-front time $x^+=0$ and in the light-front gauge $A^+=0$, up to twist-$3$ accuracy, there are four DAs, namely $\phi^{\parallel ,\perp}_V,\; g^{\perp (v,a)}_V$, for a vector meson $V$ defined via the following equations \cite{Ball:2007zt}
\begin{eqnarray}
&&\langle 0|\bar q(0)  \gamma^\mu q(x^-)| V
(P,\lambda)\rangle \nonumber \\ 
&&= f_{V} M_{V}
\frac{e_{\lambda} \cdot x}{P^+x^-}\, P^\mu
 \int_0^1 \mathrm{d} u \; e^{-iu P^+x^-}
\phi_{V}^\parallel(u,\mu)
 \label{DA:phiparallel-gvperp} \\
&&+ f_{V} M_{V}
\left(e_{\lambda}^\mu-P^\mu\frac{e_{\lambda} \cdot
	x}{P^+x^-}\right)
\int_0^1 \mathrm{d} u \; e^{-iu P^+x^-} g^{\perp (v)}_{V}(u,\mu)\nonumber  \;,
\end{eqnarray}
\begin{eqnarray}
&&\langle 0|\bar q(0) [\gamma^\mu,\gamma^\nu] q (x^-)|V
(P,\lambda)\rangle \label{DA:phiperp} \\
&& =2 f_{V}^{\perp} (e^{\mu}_{\lambda} P^{\nu} -
e^{\nu}_{\lambda} P^{\mu}) \int_0^1 \mathrm{d} u \; e^{-iuP^+ x^-} \phi_{V}^{\perp}
(u, \mu) \nnb \;,
\end{eqnarray}
and
\begin{eqnarray}
&&\langle 0|\bar q(0) \gamma^\mu \gamma^5 s(x^-)|V (P,\lambda)\rangle  \label{DA:gaperp} \\
&&=-\frac{1}{4} \epsilon^{\mu}_{\nu\rho\sigma} e_{\lambda}^{\nu}
P^{\rho} x^{\sigma}  \tilde{f}_{V} M_{V} \int_0^1 \mathrm{d} u \; e^{-iuP^+ x^-}
g_{V}^{\perp (a)}(u, \mu) \nnb \;,
\end{eqnarray}
where 
\begin{equation}
\tilde{f}_{V} = f_{V}-f_{V}^{\perp} \left(\frac{m_q + m_{\bar{q}}}{M_{K^*}} \right) \;.
\end{equation}
Note that as $x^-\to 0$, in Eqns \eqref{DA:phiparallel-gvperp} and \eqref{DA:phiperp} we recover the usual definition for the decay constant $f_V$ and $f^\perp_V$.  It follows from Eqns \eqref{DA:phiparallel-gvperp}, \eqref{DA:phiperp} and \eqref{DA:gaperp} that \cite{Ahmady:2013cva}
\begin{eqnarray}
\phi_{V}^\parallel(z,\mu) &=&\frac{N_c}{\pi f_{V} M_{V}} \int \mathrm{d}
r \mu
J_1(\mu r) [M_{V}^2 z(1-z)\nonumber \\ &+& m_{\bar{q}} m_{q} -\nabla_r^2] \frac{\phi_{L}(r,z)}{z(1-z)} \;,
\label{phiparallel-phiL}
\end{eqnarray}
\begin{equation}
\phi_{V}^\perp(z,\mu) =\frac{N_c }{\pi f_{V}^{\perp}} \int \mathrm{d}
r \mu
J_1(\mu r) [m_q - z(m_q-m_{\bar{q}})] \frac{\phi_{T}(r,z)}{z(1-z)} \;,
\label{phiperp-phiT}
\end{equation}

\begin{eqnarray}
g_{V}^{\perp(v)}(z,\mu)&=&\frac{N_c}{2 \pi f_{V} M_{V}} \int \mathrm{d} r \mu
J_1(\mu r)
\left[ (m_q - z(m_q-m_{\bar{q}}))^2 \right.\nonumber \\  &-& \left. (z^2+(1-z)^2) \nabla_r^2 \right] \frac{\phi_{T}(r,z)}{z^2 (1-z)^2
}\; ,
\label{gvperp-phiT}
\end{eqnarray}
and
\begin{eqnarray}
\frac{\mathrm{d} g_{V}^{\perp(a)}}{\mathrm{d} z}(z,\mu)&=&\frac{\sqrt{2} N_c}{\pi \tilde{f}_{V} M_{V}} \int \mathrm{d} r \mu
J_1(\mu r)
[(1-2z)(m_q^2- \nabla_r^2) \nonumber \\ &+& z^2(m_q+m_{\bar{q}})(m_q-m_{\bar{q}})]\frac{\phi_{T}(r,z)}{z^2 (1-z)^2}\; , 
\label{gaperp-phiT}
\end{eqnarray}
where $\phi_{\lambda}(r,z)$ is the light front hadronic wavefunction.


\section{Form Factors}
The seven $B\to V$ transition form factors $\{ A_0,A_1,A2,V,T_1,T_2,T_3\}$ are defined as:
{\small 
	\begin{eqnarray}
	 &&\hspace{-0.7cm}\langle V (k,\varepsilon)|\bar{q} \gamma^\mu(1-\gamma^5 )b | B(p) \rangle \nonumber \\
	&&= \frac{2i V(q^2)}{m_B + m_{V}} \epsilon^{\mu \nu \rho \sigma} \varepsilon^*_{\nu} k_{\rho} p_{\sigma} -2m_{K^*} A_0(q^2) \frac{\varepsilon^* \cdot q}{q^2} q^{\mu}  \nonumber \\
	&&- (m_B + m_{V}) A_1(q^2) \left(\varepsilon^{\mu *}- \frac{\varepsilon^* \cdot q q^{\mu}}{q^2} \right) \nonumber \\
	&&+ A_2(q^2) \frac{\varepsilon^* \cdot q}{m_B + m_{V}}  \left[ (p+k)^{\mu} - \frac{m_B^2 - m_{V}^2}{q^2} q^{\mu} \right]
	\end{eqnarray}
	\begin{eqnarray}
	&&\hspace{-0.7cm}q_{\nu} \langle V (k,\varepsilon)|\bar{d} \sigma^{\mu \nu} (1-\gamma^5 )b | B(p) \rangle \nonumber \\ &&= 2 T_1(q^2) \epsilon^{\mu \nu \rho \sigma} \varepsilon^*_{\nu} p_{\rho} k_{\sigma} \nonumber \\
	&&- i T_2(q^2)[(\varepsilon^* \cdot q)(p+k)_{\mu}-\varepsilon_{\mu}^*(m_B^2-m_{V}^2)] \nonumber \\
	&&- iT_3(q^2) (\varepsilon^* \cdot q) \left[ \frac{q^2}{m_B^2-m_{V}^2} (p+k)_{\mu} -q_{\mu}  \right] 
	\end{eqnarray}
} 
We use LCSR\cite{ali,ball1,ball2,ball3,aliev} to calculate the seven form factors with AdS/QCD DAs.  For example, the LCSR for the radiative form factor $T_1$ is given below:
\begin{eqnarray}
&&\hspace{-0.7cm}T_1(q^2) = \frac{1}{4} \left( \frac{m_b}{f_B m_B^2}\right) \exp{\left(\frac{m_B^2}{M^2}\right)}
\int_\delta^1 \frac{\d u}{u}\, \nnb \\&& \exp \ga - \frac{m_b^2 + p^2 u \bar u - q^2 \bar u}{uM^2} \dr \Bigg\{ m_b f_V^\perp \phi_\perp (u) + \nnb \\&&
\; f_V m_V \Bigg[ \Phi_\parallel (u)  +
u g_\perp^{(v)} (u)  +\frac{g_\perp^{(a)}(u)}{4} \nnb \\&& + \frac{(m_b^2 + q^2 -p^2 u^2)g_\perp^a (u)}{4 u M^2} \Bigg] \Bigg\}
\label{LCSRT1}
\end{eqnarray}
In Eqn. \eqref{LCSRT1}, $M$ is the Borel parameter and $\delta$ is associated with the continuum threshold \cite{ali}.  


\section{Holographic Light Front Wavefunction}
It remains now to specify the light front wavefunction for the light vector mesons.  In light-front QCD, with massless quarks, the meson wavefunction can be written in the following factorized form\cite{Brodsky}:
\begin{equation}
\phi(z,\zeta, \varphi)=\frac{\Phi(\zeta)}{\sqrt{2\pi \zeta}} f(z) \mathrm{e}^{i L \varphi}  
\label{factorized-lc}
\end{equation}
with $\Phi(\zeta)$ satisfying the so-called holographic light-front Schr\"{o}edinger equation 
\begin{equation}
\left(-\frac{d^{2}}{d\zeta^2}-\frac{1-4L^{2}}{4\zeta^{2}}+U(\zeta)\right)\Phi(\zeta)=M^{2}\Phi(\zeta)
\label{hLFSE} 
\end{equation}
where $L$ is the orbital angular momentum quantum number, $z$ is the fraction of the meson light-front momentum carried by the quark and the variable $\zeta=\sqrt{z(1-z)}r$ where $r$ is the transverse distance between the quark and the antiquark. The AdS/QCD correspondence dictates that $f(z)=\sqrt{z(1-z)}$\cite{Brodsky}. The confining potential
\begin{equation}
U(\zeta)=\kappa^4 \zeta^2 + 2\kappa^2(J-1) 
\label{quadratic-dilaton}
\end{equation}
is obtained either from soft wall model in AdS or from considerations of one-dimensional conformal field theory\cite{physicsreport}.  The holographic light-front wavefunction for a vector meson $(L=0,S=1)$ then becomes
\begin{eqnarray}
\phi_{\lambda} (z,\zeta) \propto \sqrt{z(1-z)} \exp
\left(-\frac{\kappa^2 \zeta^2}{2}\right)\nnb \\
\times\exp\left \{-\left[\frac{m_q^2-z(m_q^2-m^2_{\bar{q}})}{2\kappa^2 z (1-z)} \right] 
\right \} \label{AdS-QCD-wfn}
\end{eqnarray}
with $\kappa=M_{V}/\sqrt{2}$ and where we have introduced the dependence on quark masses following a prescription by Brodsky and de T\'eramond \cite{Brodsky2}.  This wavefunction for the $\rho$ meson was successfully used to predict the diffractive $\rho$ meson electroproduction at HERA\cite{PRL}.
\section{Results}
In Ref. \cite{Ahmady:2013cva}, we have used the $\rho$ DAs to compute the branching ratio for the radiative decay $B \to \rho \gamma$ including the power suppressed corrections.  In the same work, we calculated the branching ratio for the very rare power-suppressed decay $B_s \to \rho \gamma$. Furthermore, in Ref. \cite{PRD2}, we have used the DAs for the $K^*$ to compute the isospin asymmetry in $B \to K^* \gamma$ decay.  For $\bar B^\circ\rightarrow \rho^\circ \gamma$, our predictions agree with those generated using Sum Rules DAs and with the data from the BaBar and Belle collaborations. In computing the weak annihilation amplitude which is power-supressed in $\bar B^\circ\rightarrow \rho^\circ \gamma$ but is the leading contribution in $\bar{B}^{\circ}_s\rightarrow \rho^\circ \gamma$, we find that the AdS/QCD DA avoids the end-point divergences encountered with the SR DA.  Our prediction for the branching ratio agrees with that obtained using SR DAs and with experiment. More interestingly, our prediction for the isospin asymmetry using the AdS/QCD DA  does not suffer from the end-point divergence encountered when using the corresponding SR DA.  

We also use the AdS/QCD DAs to compute the transition form factors for $B\to (\rho ,\, K^*)$ decays using LCSR\cite{PRD3,PRD4}.  In Ref. \cite{PRD3}, we obtained the three form factors $A_0$, $A_1$, $A_2$ and $V$ which are relevant for the semileptonic $B\to\rho\ell\nu$ decay.  Having computed the form factors, we are able to compute $V_{ub}$-independent ratios of partial decay widths in various $q^2$ bins and compare it with the available experimental measurement\cite{BaBar}
\begin{equation}
\Gamma_{\mbox{\tiny{low}}}= \int_0^8 \frac{\mathrm{d} \Gamma}{\mathrm{d} q^2} \mathrm{d} q^2 = (0.564 \pm 0.166) \times 10^{-4} 
\end{equation}
for the low $q^2$ bin,
\begin{equation}
\Gamma_{\mbox{\tiny{mid}}}= \int_8^{16} \frac{\mathrm{d} \Gamma}{\mathrm{d} q^2} \mathrm{d} q^2 = (0.912 \pm 0.147) \times 10^{-4} 
\end{equation}
for the intermediate $q^2$ bin and
\begin{equation}
\Gamma_{\mbox{\tiny{high}}}= \int_{16}^{20.3} \frac{\mathrm{d} \Gamma}{\mathrm{d} q^2} \mathrm{d} q^2 = (0.268 \pm 0.062) \times 10^{-4} 
\end{equation}
for the high $q^2$ bin. 
From these measurements, we can thus deduce the $|V_{ub}|$-independent ratios of partial decay widths
\begin{equation}
R_{\mbox{\tiny{low}}}=\frac{\Gamma_{\mbox{\tiny{low}}}}{\Gamma_{\mbox{\tiny{mid}}}}=0.618 \pm 0.207
\end{equation}
and 
\begin{equation}
R_{\mbox{\tiny{high}}}=\frac{\Gamma_{\mbox{\tiny{high}}}}{\Gamma_{\mbox{\tiny{mid}}}}=0.294 \pm 0.083
\end{equation}
which we compare to our predictions: $R_{\mbox{\tiny{low}}}=0.580, 0.424$, $R_{\mbox{\tiny{high}}}=0.427,0.503$ for $m_q=0.14, 0.35$ GeV respectively.  Our predictions for $R_{\mbox{\tiny{low}}}$ are therefore in agreement with the BaBar measurement. This is not the case for $R_{\mbox{\tiny{high}}}$ where our predictions are above the BaBar measurement. This is perhaps not unexpected given that the LCSR predictions are less reliable in the high $q^2$ bin. 

\begin{figure}
	\vspace*{-1.75cm}\includegraphics[width=0.50\textwidth]{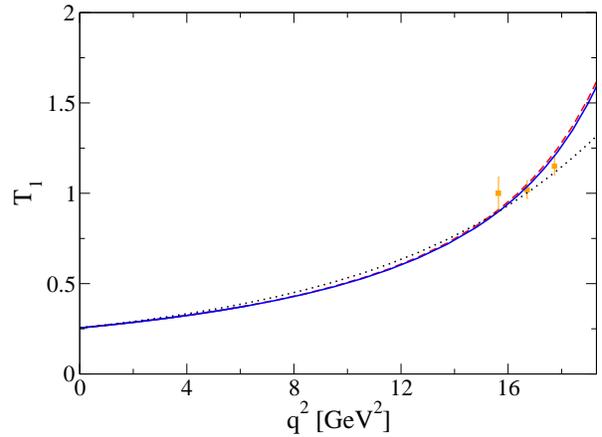}\\
	\vspace*{-0.5cm}\includegraphics[width=0.50\textwidth]{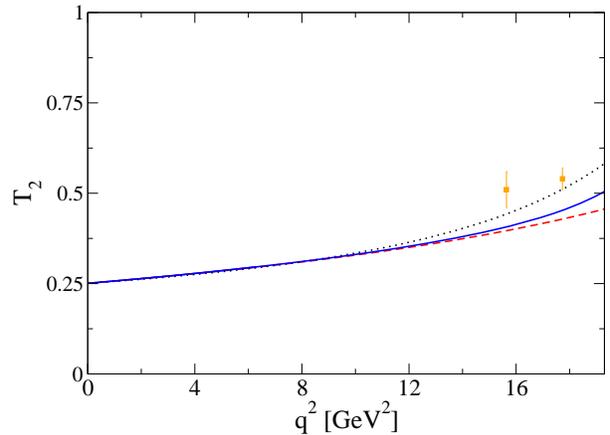}
	\caption{Two of the seven $B\to K^*$ transition form factors. The lattice data is from Ref. \cite{latticekstar}. AdS/QCD: solid blue, AdS/QCD $+$ lattice : dotted black. AdS/QCD fit: dashed red.}
	\label{fig:FFKstar}       
\end{figure}

For the $B \to K^*$ transition, there are seven relevant form factors. Again we compute them using LCSR with the $K^*$ AdS/QCD holographic DAs as input. Our results for two of the three tensor form factors $T_1$ and $T_2$ are shown in Figure \ref{fig:FFKstar}. The results for the full set of form factors can be found in Ref. \cite{PRD4}. The solid blue curves are the AdS/QCD-LCSR predictions extrapolated to high $q^2$. We also do  fits using the form
\begin{equation}
F(q^2)=\frac{F(0)}{1- a (q^2/m_B^2) + b (q^4/m_B^4)}
\label{FitFF}
\end{equation}
to the AdS/QCD predictions in the region of reliability of the LCSR and extrapolate to high $q^2$. These are the red dashed curves. Finally, we include in our fits the lattice data \cite{latticekstar} available at large $q^2$ to generate the dashed black curves. Our results for the above fits are given in Table \ref{tab:abAdS}. 

\begin{table}[t]
	\caption{The values of the form factors at $q^2=0$ together with the fitted parameters $a$ and $b$. The values of $a$ and $b$ are obtained by fitting Eq. \eqref{FitFF} to either the AdS/QCD predictions for low-to-intermediate $q^2$ or both the AdS/QCD predictions for low-to-intermediate $q^2$ and the lattice data at high $q^2$.}
	\centering
	\label{tab:abAdS}       
	\begin{tabular}{llllll}
		\hline\noalign{\smallskip}
		& & \multicolumn{2}{l}{AdS/QCD} & \multicolumn{2}{l}{AdS/QCD + Lat.} \\ \cline{3-4} \cline{5-6}
		F & $F(0)$ & $a$ & $b$ & $a$ & $b$ \\[3pt]
		$A_0$ & $0.285$ & $1.158$ & $0.096$ & $1.314$ & $0.160$ \\
		$A_1$ & $0.249$ & $0.625$ & $-0.119$ & $0.537$ & $-0.403$ \\
		$A_2$ & $0.235$ & $1.438$ & $0.554$ & $1.895$ & $1.453$ \\
		$V$ & $0.277$ & $1.642$ & $0.600$ & $1.783$ & $0.840$ \\
		$T_1$ & $0.255$ & $1.557$ & $0.499$ & $1.750$ & $0.842$ \\
		$T_2$ & $0.251$ & $0.665$ & $-0.028$ & $0.555$ & $-0.379$  \\
		$T_3$ & $0.155$ & $1.503$ & $0.695$ & $1.208$ & $-0.030$ \\
		\noalign{\smallskip}\hline
	\end{tabular}
\end{table}

\begin{figure}
	\centering
	\includegraphics[width=0.5\textwidth]{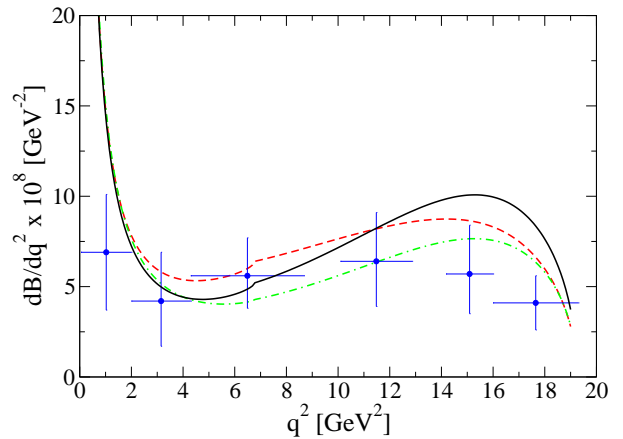}
	\caption{The differential branching ratio of the $B \to K^* \mu^+ \mu^-$ process as a function of $q^2$. The dashed red line is the AdS/QCD result, solid black is the AdS/QCD+lattice fit and the dot-dash green curve is the AdS/QCD+lattice+NP result. Note that we average the LHCb experimental result data from both the $B^{\circ} \to K^{*\circ} \mu^+ \mu^-$ \cite{LHCb} and $B^{+} \to K^{*+} \mu^+ \mu^-$ \cite{LHCb2} data.}
	\label{fig:BR}       
\end{figure}

Finally, we use the $B \to K^*$ form factors to compute the differential branching ratio for $B \to K^* \mu^+ \mu^-$. Our results are shown in figure \ref{fig:BR}. As can be seen, our AdS/QCD prediction (dashed red curve) tend to overshoot the data at high $q^2$. Using the form factors fitted to the lattice data does not remedy the situation: see the black solid curve. On the other hand, we are able to achieve agreement at high $q^2$ by adding a new physics contribution to the Wilson coefficient $C_9$ \cite{Descotes-Genon:2013vna}.   

%


\end{document}